\documentclass[11pt,a4paper,reqno]{amsart}
\usepackage{amsmath}
\usepackage[english]{babel}
\usepackage{latexsym}
\usepackage{amssymb}
\usepackage{amscd}
\usepackage{amsgen,amstext,amsbsy,amsopn}
\usepackage{math rsfs}
\usepackage{bm,bbm}
\usepackage{amsthm,epsfig,graphicx,graphics}
\usepackage[latin1]{inputenc}
\usepackage{xspace}
\usepackage{amsxtra}
\usepackage{xcolor}
\usepackage{dsfont}
\usepackage{enumerate}

\usepackage{geometry}
\newtheorem{theorem}{Theorem}[section]

\newtheorem{proposition}{Proposition}[section]
\newtheorem{corollary}{Corollary}[section]

\newcommand{\disp}{\displaystyle}
\newcommand{\tx}{\textstyle}

\newcommand{\bdm}{\begin{displaymath}}
\newcommand{\edm}{\end{displaymath}}
\newcommand{\bdn}{\begin{eqnarray}}
\newcommand{\edn}{\end{eqnarray}}
\newcommand{\bay}{\begin{array}{c}}
\newcommand{\eay}{\end{array}}
\newcommand{\ben}{\begin{enumerate}}
\newcommand{\een}{\end{enumerate}}
\newcommand{\beq}{\begin{equation}}
\newcommand{\eeq}{\end{equation}}
\newcommand{\beqn}{\begin{eqnarray}}
\newcommand{\eeqn}{\end{eqnarray}}

\newcommand{\lf}{\left}
\newcommand{\ri}{\right}

\newcommand{\ket}[1]{\lf|#1 \ri\rangle}

\newcommand{\kett}[1]{\lf|#1 \ri)}

\newcommand{\R}{\mathbb{R}}

\newcommand{\C}{\mathbb{C}}

\newcommand{\Z}{\mathbb{Z}}

\newcommand{\xv}{\mathbf{x}}
\newcommand{\yv}{\mathbf{y}}

\newcommand{\kv}{\mathbf{k}}

\newcommand{\nv}{\mathbf{n}}

\newcommand{\diff}{\mathrm{d}}

\newcommand{\eps}{\varepsilon}

\renewcommand{\L}{\Lambda}

\newcommand{\tr}{\mathrm{Tr}}

\newcommand{\hamb}{\mathcal{H}}

\newcommand{\kb}{\mathcal{K}}

\newcommand{\spin}{\hat{S}}

\newcommand{\spinv}{\hat{\mathbf{S}}}

\newcommand{\up}{a^{\dagger}}
\newcommand{\ax}{a_{\xv}}
\newcommand{\ay}{a_{\yv}}

\newcommand{\upx}{\up_{\xv}}
\newcommand{\upy}{\up_{\yv}}

\newcommand{\nx}{n_{\xv}}

\newcommand{\hnx}{\hat{n}_{\xv}}
\newcommand{\hny}{\hat{n}_{\yv}}

\newcommand{\hH}{\mathscr{H}}

\newcommand{\F}{\mathcal{F}}

\begin{document}

\title{Low-Temperature Spin-Wave Approximation for the Heisenberg Ferromagnet}

\thanks{Contribution to the proceedings of ICMP2015, Santiago de Chile, July 27 -- August 1, 2015.}

\author{M. Correggi}
\author{A. Giuliani}

\address{Dipartimento di Matematica e Fisica, Universit\`{a} degli Studi Roma Tre,\\
Rome, 00146, Italy}

\author{R. Seiringer}

\address{Institute of Science and Technology Austria (IST Austria), Klosterneuburg, 3400, Austria}

\begin{abstract}
We study the low temperature thermodynamics of the quantum  Heisenberg ferromagnet in three dimension for any value of the spin $ S \geq 1/2 $. We report on a rigorous proof of the validity of the spin-wave approximation for the excitation spectrum, at the level of the first non-trivial contribution to the free energy as the inverse temperature $ \beta \to \infty $. 	
\end{abstract}


\maketitle

\section{Introduction}

The Heisenberg Model (HM) played a special role in statistical mechanics as a paradigm of discrete interacting systems with continuous symmetry, but, at the same time, provided a simple but reliable model for magnetism in many materials. Its importance in both the mathematical and physical literature can not be overlooked: among the major achievements on the HM, we recall the celebrated Mermin-Wagner theorem  \cite{FP,H,MW} about the absence of phase transitions in low dimensions and the rigorous proofs of symmetry breaking for the classical  \cite{FSS} and quantum anti-ferromagnetic \cite{DLS} HM. 

The investigation of the low-temperature properties of the isotropic HM is typically performed within an approximation -- the {\it spin-wave} theory -- which dates back to the `30s  \cite{Bl} and which amounts to restricting the analysis to long-range excitations -- the spin-waves -- of a selected ground state. Among the successful predictions of spin-wave theory we recall the low temperature behavior \cite{Bl} of the spontaneous magnetization , later verified experimentally, and the occurrence of a phase transition associated to breaking of the rotational symmetry at low temperature in three or more dimensions. Spin-wave theory was also involved in the study \cite{Dy1,Dy2} by Dyson of the low temperature expansion of the magnetization, which solved a long standing and debated question about the first non-trivial contribution of the spin-wave interaction. Nowadays spin-wave theory is still widely used in  physics to extract reliable predictions on several critical phenomena, e.g., Bose-Einstein condensation in magnets \cite{BNY,Y}. 

Concerning the mathematical point of view the grounds of spin-wave theory has been investigated in several cases: among the most relevant contributions we mention the proof of the exactness of spin-wave expansion to any order in the classical $N-$vector HM \cite{Ba1,Ba2,Ba3} and in the plane rotator \cite{BFLLS} models. The ferromagnetic HM is notably missing in this list of results, namely neither the existence of a broken phase, nor the exactness of the spin-wave theory have been proven. We report here on some recent progress \cite{CGS1,CGS2} in this latter direction. 

To our knowledge the best results available to date were the upper bounds to the free energy in the low temperature regime proven by Conlon-Solovej \cite{CS2} and Toth \cite{T}: the latter result improved the first one but, although both of the them captured the correct temperature power law $ T^{5/2} $, the bounds were not sharp in the constant prefactor (see also below). It is also worth mentioning that in the limit of large spin more results are actually available both in the classical \cite{L} and in the quantum regime \cite{CG}.

Our main result, stated in Sect. \ref{sec: main results}, is the asymptotic {\it exactness of the spin-wave theory} for the computation of the free energy of the quantum Heisenberg ferromagnet {\it as $ \beta \to \infty$}, with $ \beta $ inverse temperature, and {\it for any spin $ S > 0 $}. An interesting consequence of this result is an instance of long-range order: combining the free energy asymptotics with a operator bound on the Hamiltonian, we can show that the two-point function is separated from zero up to the length scale $ \beta^{5/4} $.

In the next Sect. \ref{sec: model} we discuss the details of the model and introduce the spin-wave approximation from both the heuristic and rigorous points of view. After the statement of the main results in Sect. \ref{sec: main results}, we present a brief sketch of the main steps in the proofs.

\section{Ferromagnetic Heisenberg Model} 
\label{sec: model}

In order to keep the setting as simple as possible, we consider the three-dimensional ferromagnetic HM with nearest neighbor interaction, i.e., the quantum Hamiltonian
\begin{equation}
	\label{eq: ham}
	H =  \sum_{\langle \xv,\yv \rangle \subset \Lambda} \lf(S^2 - \spinv_{\bf x} \cdot \spinv_{\bf y} \ri),
\end{equation}
where $\Lambda\subset \Z^3$ is a cube with $ L $ sites per side, the sum is over all (unordered) nearest neighbor pairs $\langle \xv,\yv\rangle$ in $\Lambda$, and $ \spinv_{\xv} $ is a spin $ S $ operator. The associated specific free energy in $\L$ is given by 
\beq 
	f(S,\beta,\Lambda)=-\frac1{\beta L^3}\ln \tr e^{-\beta H_\Lambda}
\eeq
and its thermodynamics limit is denoted by $f(S,\beta)$. We assume free conditions on the boundary of $ \L $ although our result will be clearly independent of such a choice. The Hilbert space is $ \hH = \C^{(2S+1) L^3} $. 

The additive constant $S^2$ in \eqref{eq: ham} normalizes the ground state energy of $H$ to zero. Indeed the ground state energy of the operator \eqref{eq: ham} is reached on states with maximal total spin $ S_T = SL^3 $, where
\beq
	\spinv_T : = \sum_{\xv \in \L} \spinv_{\xv},
\eeq
and the eigenvalues of $ \spinv_T^2 $ are given by $ S_T(S_T + 1) $. This easily follows from the fact that states with maximal $ S_T $ are the only ones such that any partial sum of spins is maximal as well, i.e., given $ N < L^3 $ points $ \xv_i, \ldots, \xv_N \in \L $, $ (\spinv_{\xv_1} + \cdots + \spinv_{\xv_N})^2 $ equals $ N S (N S + 1) $ on such states. In particular this holds true for any nearest neighbor sites $ \langle \xv,\yv \rangle $ and therefore $ \langle \spinv_{\xv} \cdot \spinv_{\yv} \rangle = S^2 $ on states with $S_T = S L^3 $.

Spin-waves can be naturally introduced as excited states of $ H $, with respect to a given ground state. Because of the rotational invariance of the model the ground state is indeed highly degenerate, but if we pick one of such states, e.g., the one with the third components of all spins being equal to the maximal value $ +S $, which we denote by $ \bigotimes_{\xv \in \L} \ket{S_{\xv}^3 = S} $, then the spin-wave $ \ket{\kv} $ with momentum $ \kv \in \Lambda^* $, i.e., $ \kv = \frac{2\pi}{L} \nv $, $ \nv = (n_1, n_2, n_3) $, $ n_i = 0, \ldots, L-1 $, is
\beq
	\ket{\kv} = \frac{1}{L^{3/2}} \sum_{\xv \in \L} e^{-i \kv \cdot \xv} \spin_{\xv}^{-}  \bigg[ \bigotimes_{\yv \in \L} \ket{S_{\yv}^3 = S} \bigg],
\eeq
where $ \spin_{\xv}^{\pm} : = \spin_{\xv}^1 \pm i \spin_{\xv}^2 $. A spin-wave state is thus the quantum analogue of a classical spin wave, in which all the spins are deflected by a small angle in a coherent way. Spin-wave states are orthonormal eigenstates of $ H $ with eigenvalues $ S \eps(\kv) $, where
\beq
	\eps(\kv) = 2 \sum_{i=1}^3(1-\cos k_i),
\eeq
is the typical dispersion relation of particles on a lattice.

Heuristically a spin-wave expansion can be given in terms of states containing one or more spin-waves. Such states are however not orthogonal, which makes the expansion only formal. An alternative way to introduce spin-waves is via the Holstein-Primakoff \cite{HP} (HP) representation, which maps $ H $ onto a bosonic operator acting on a suitable Fock space: by setting
\begin{equation}
	\label{eq: ax upx}
	\spinv_{\xv}^+ = :  \sqrt{2S}\, \upx \sqrt{1 - \tx\frac{\upx \ax}{2S} },		\quad	\spinv_{\xv}^{-} = : \sqrt{2S} \sqrt{ 1 - \tx\frac{\upx \ax}{2S} } \ax,		\quad	\spinv_{\xv}^3 = : \upx \ax - S,
\end{equation}
one can identify a pair of creation and annihilation operators per site $ \xv \in \L $ satisfying the canonical commutation relations $ [\ax,a^{\dagger}_{\yv}] = \delta_{\xv,\yv} $. A basis of the Fock space $ \F_S $ is given by states $ \bigotimes_{\xv \in \L} \kett{\nx} $, with $ n_{\xv} = 0, \ldots, 2S $ bosons at site $ \xv $. The correspondence with spin states is given by 
\bdm
	\kett{\nx} \longleftrightarrow  \ket{S^3_{\xv} = \nx - S},
\edm
which is obviously a basis of the single-site Hilbert space $ \C^{2S+1} $. Notice the presence of the hard-core constraint $ \nx \leq 2S $ in $ \F_S $, i.e., $ \upx \kett{\nx = 2S} = 0 $ by definition, which is also crucial for the well-posedness of \eqref{eq: ax upx}.

Under this transformation the Hamiltonian $ H $ is mapped onto the operator
\beq
	\label{eq: hamb}
	\hamb = \hamb_0 + \kb
\eeq
with 
\beq
	\hamb_{0}  =   S \disp\sum_{\langle \xv,\yv \rangle \subset \Lambda} \lf(\upx - \upy \ri) \lf( \ax - a_{\yv} \ri),
\eeq
the free part corresponding to the second quantization of the discrete Laplacian on $ \L $ and an interaction
\beq
	\kb = \disp\sum_{\langle \xv,\yv \rangle \subset \Lambda} 
	\lf\{- \upx \up_{\yv} \ax a_{\yv} + 2S \upx \bigg[1 - \sqrt{ 1 - \tx\frac{\hnx}{2S}} \sqrt{ 1 - \tx\frac{\hny}{2S} } \bigg] a_{\yv} \ri\},
\eeq
which is mostly attractive and contains terms of higher order than the quartic one, i.e., it is not only two-body.

The spin-wave approximation can now be made more explicit: at leading order it amounts of dropping both the interaction $ \kb $ {\it and} the hard-core constraint\footnote{This of course has to be taken with care for the $ \kv = 0 $ mode.} $ \nx \leq 2S $. The result is a bosonic gas of non-interacting particles, whose free energy in the thermodynamic limit can be explicitly computed and converges to
\bdm
	f_0(\beta,S) = \disp\frac{1}{(2\pi)^3\beta} \int_{[-\pi,\pi]^3} \diff\kv \: \log \lf( 1 -e^{ - \beta  S  \eps(\kv)} \ri).
\edm
In the low temperature limit, by scaling the momenta, we thus obtain
\beq
	\label{eq: f0}
 	f_0(\beta,S) = C_0 S^{-3/2} \beta^{-5/2} (1 + o(1)),	
\eeq
\beq
	\label{eq: c0}
	C_0 : = \frac{1}{(2\pi)^3} \int_{\R^3} \diff\kv \: \log \lf( 1 -e^{ - k^2} \ri) = - \frac{\zeta(5/2)}{8 \pi^{3/2}},
\eeq
with $ \zeta $ the Riemann zeta function. In our main Theorem we prove {\it both} upper and lower bounds to the free energy $ f(\beta,S) $ in terms of $ f_0(\beta,S) $, which are sharp in the $ \beta \to \infty $ limit.

\section{Main Results}
\label{sec: main results}

The expression \eqref{eq: f0} is therefore the one we expect to recover expanding $ f(\beta,S) $ to leading order in $ \beta $, as $ \beta \to \infty $. Before stating our main result we comment at this stage on older results \cite{CS2,T}: the upper bounds proven there have the form (both results were proven for $ S = 1/2 $)
\bdm
	f(\beta,S) \leq C \: 2^{3/2} \beta^{-5/2},
\edm
for some $ C > C_0 $. The best constant \cite{T} was $ C = C_0 \log 2 $. More results \cite{CGS1,CS1} are available in the regime $ S \to \infty $, with $ \beta \propto S^{-1} $.

\begin{theorem}[Free energy asymptotics]
	\label{teo: main}
	\mbox{}	\\
	For any $ S \geq 1/2 $,
	\beq
		\label{eq: main}
		\lim_{\beta \to \infty} S^{3/2} \beta^{5/2} f(\beta,S) = C_0,
	\eeq
	with $C_0 $ given by \eqref{eq: c0}.
\end{theorem}

The result is actually uniform in $ S $, for $ S < + \infty $, and the proof applies to the case $ S \to \infty $ too, but under the additional assumption $ \beta S \gg \log S $. So in this respect it can be thought of the convergence of the free energy whenever $ \beta S \to \infty $, with $ \beta S \gg \log S $. Also the proof can be easily generalized to any dimension larger than three, but for concreteness we stick to the physically more significant setting.

As anticipated, a consequence of the free energy asymptotics is the persistence of spin order up to length scales of order $ \beta^{5/4} $ as $ \beta \to \infty $.

\begin{corollary}[Quasi long-range order]
	\label{cor: lro}
	\mbox{}	\\
	For any $ \xv, \yv \in \L $, with $ |\xv - \yv| \ll \beta^{5/4} $,
	\beq
		\label{eq: lro}
		\langle \spinv_{\xv} \cdot \spinv_{\yv} \rangle_{\beta} \geq S^2 - C \beta^{-5/2} |\xv - \yv |^2 = S^2 + o(1) > 0,
	\eeq
	where the expectation value is on any translational invariant Gibbs state in infinite volume with inverse temperature $ \beta \gg 1 $.
\end{corollary}

\section{Sketch of the Proofs}
\label{sec: proofs}

Here we briefly sketch the main steps of the proof of Theorem \ref{teo: main} and Corollary \ref{cor: lro}, which is given elsewhere in full details \cite{CGS2} or in the simpler case \cite{CGS1} $ S = 1/2 $.

Theorem \ref{teo: main} is proven by comparing suitable upper and lower bounds to the free energy density $ f(\beta, S) $ in the thermodynamic limit.

The upper bound part of the proof is as usual the easiest one and relies on the Gibbs variational principle $ f(S,\beta,\Lambda) \leq  \frac{1}{|\Lambda|} \tr H_\Lambda \Gamma  + \frac{1}{\beta |\Lambda|} \tr  \Gamma \ln \Gamma $, which requires only to provide a trial state to compute the upper bound. The preliminary step is however the confinement of particles into boxes of side length $  \gg \sqrt{\beta} $ with Dirichlet boundary conditions.

The trial state is then the closest possible one to a Gibbs state of $ \hamb_0 $, i.e.,
\beq
	\Gamma = \frac{ P e^{-\beta  \hamb_0} P }{\tr_{\F_S} P e^{-\beta  \hamb_0}},
\eeq
where $ P $ denotes the hard-core projection onto states with at most one particle per site. The key step in the upper bound is the replacement of such projection, which is made by exploiting the simple operator inequality 
\bdm
	1 - P \leq \sum_{\xv \in L} \hnx \lf(\hnx - 1 \ri),
\edm
and then implementing Wick's rule.

The proof of the lower bound is much more involved and is done in several steps:
\begin{enumerate}
	\item we first localize into Neumann boxes;
	\item the we prove a preliminary lower bound to the free energy (off by a factor $ \log\beta $) by means of a sharp spectral estimate on $ H $;
	\item exploiting the rotational invariance, we estimate the trace via the HP mapping;
	\item we finally use Peierls-Bogoliubov inequality and bound the interaction expectation value.
\end{enumerate}

Step (1) is essentially trivial because it just suffices to divide the box into smaller ones and drop the links between different boxes, since the contribution of any nearest neighbor pair to $ H $ is positive. 

The preliminary bound on $ f(\beta,S) $ at Step (2) is a consequence of the following operator bound on the excitation spectrum of $ H_{\ell} $, the Heisenberg Hamiltonian on a box of length $ \ell $ with Neumann boundary conditions.

\begin{proposition}[Excitation spectrum bound]
	\label{pro: bound h}
	\mbox{}	\\
	There exists a positive constant $ C > 0 $ such that
	\begin{equation}\label{ham lb}
		\lf.H_{\ell}\ri|_{\hH_{S_T}} \ \geq \frac{CS}{\ell^{2}} \left( S \ell^3 - S^T \right),
	\end{equation}
	where $ \hH_{S_T} $ is the subspace of $ \hH $ with $ \spinv_T^2 = S_T(S_T+1)$.
\end{proposition}

Except for the constant the bound is sharp, in particular for the energy gap above the ground state energy, which is estimated as $ CS \ell^{-2} $, and allows to show that, in a box $ \Lambda_{\ell} $ with side length $ \ell $ and Neumann boundary conditions
\beq
	\label{eq: preliminary}
	f(S,\beta,\Lambda_\ell) \geq - C S  \left[ \ln( S\beta) / ( S\beta) \right]^{5/2},
\eeq
provided $ \ell \geq \sqrt{\beta S} $ and $ \beta S \gg \log S $.

By means of the above bound we are then able to discard from the computation of the partition function states with energy larger than $ C \ell^3 \beta^{-5/2+\epsilon} $. In this subspace we pass to the bosonic HP representation (Step (3)), resolving the degeneracy due to rotational invariance by selecting states with total spin $ S_T $ and third component equal to $ - S_T $. For such states the restriction induced by \eqref{eq: preliminary} becomes a bound on the total number of bosonic particles, which can not exceed $ C \ell^5 \beta^{-5/2+\epsilon} $. The rotational symmetry degeneracy generates an unessential prefactor $ 2S_T + 1 $.

We then consider boxes of side length slightly larger than $ \sqrt{\beta} $, so that we are left with a very dilute gas of interacting bosons. To conclude the proof (Step (4)) it just remains to apply Peierls-Bogoliubov inequality and estimate the expectation value of the interaction onto eigenstates of the full bosonic Hamiltonian. This is done by exploiting a pointwise bound on the two-point function proven in next Theorem: we first show that $ \kb \leq \frac 12 \sum_{\langle \xv,\yv \rangle}  \left( 4 \hnx \hny + \hnx (\hnx -1) +  \hny( \hny -1) \right) $, so that the expectation value of $ \kb $ in an eigenstate $ \Psi $ of $ \hamb $ is bounded as $ \langle \Psi | K | \Psi\rangle \leq 18 \ell^3 \lf\|\rho \ri\|_{\infty} $, with $ \rho $ the two-particle density, i.e.,
\begin{equation}
	\rho(\xv,\yv) : = \langle\Psi | \upx \upy \ax \ay | \Psi \rangle.
\end{equation}

\begin{proposition}[Interaction estimate]
	\label{pro: rho}
	\mbox{}	\\
	There exists a constant $C>0$ such that, if $\Psi$ is an eigenfunction of $ \hamb $ on $\Lambda_
\ell $ with energy $E>0$, then
	\begin{equation}\label{thm:eq:rho}
		\lf\|\rho \ri\|_{L^\infty(\Lambda_{\ell} \times \Lambda_{\ell})} \leq C  S^{-3} E^{3} \lf\|\rho \ri\|_{L^1(\Lambda_{\ell} \times \Lambda_{\ell})}.
	\end{equation}
\end{proposition}

The proof of the above result, which might be of a certain interest in itself, relies on a differential inequality for $ \rho $, allowing to reduce the $ N-$body problem to a two-body one, and a iteration of such inequality to estimate $ \lf\|\rho \ri\|_{\infty}$ in terms of the asymptotic probability density of a modified random walk on $ \Z^6 $.

To complete the proof of Theorem \ref{teo: main}, we combine Proposition \ref{pro: rho} with the a priori bounds on the energy $ E $ and the particle number, obtaining 
\beq
	\langle \Psi | K | \Psi\rangle \leq C_S \ell^7 E^5 \lesssim \beta^{-3/2+\epsilon}.
\eeq
where we have chosen $ \ell \simeq \beta^{1/2 + \epsilon} $. The above quantity is in turn much smaller that the expectation value of $ \hamb_0 $, which is of order $ \ell^{-2} \simeq \beta^{-1+\epsilon} $ and therefore contributes only at higher order.

A variant of the proof of Proposition \ref{pro: bound h} yields the inequality
\bdm
	\langle S^2 - \spinv_{\xv} \cdot \spinv_{\yv} \rangle_{\beta} \leq C  |\xv - \yv|^2 e(\beta, S),
\edm
with $ e(\beta,S) = \partial_{\beta}(\beta f(\beta,S)) $ the energy per site. Hence Theorem \ref{teo: main} directly implies Corollary \ref{cor: lro}.

\bigskip
\noindent
{\bf Acknowledgements.} M.C. acknowledges the support of MIUR through the FIR2013 grant ``Cond-Math" (code RBFR13WAET). R.S. acknowledges the support by the Austrian Science Fund (FWF), project Nr. P27533-N27.

\end{document}